\documentclass[prd,a4paper]{revtex4}
\usepackage{epsfig}
\newcommand{\be}{\begin{equation}}
\newcommand{\ee}{\end{equation}}
\newcommand{\bea}{\begin{eqnarray}}
\newcommand{\eea}{\end{eqnarray}}
\newcommand{\bean}{\begin{eqnarray*}}
\newcommand{\eean}{\end{eqnarray*}}

\newcommand{\gapproxeq}{\lower
.7ex\hbox{$\;\stackrel{\textstyle >}{\sim}\;$}}
\newcommand{\lapproxeq}{\lower
.7ex\hbox{$\;\stackrel{\textstyle <}{\sim}\;$}}

\begin{document}

\bibliographystyle{unsrt}

\title{\bf Hadronic loop contributions to $J/\psi$ and $\psi^\prime$ radiative decays into $\gamma\eta_c$
or $\gamma\eta_c^\prime$}

\author{Gang Li$^{1,3}$ and Qiang Zhao$^{2,1,3}$}

\affiliation{1) Institute of High Energy Physics, Chinese Academy
of Sciences, Beijing 100049, P.R. China}

\affiliation{2) Department of Physics, University of Surrey,
Guildford, GU2 7XH, United Kingdom}

\affiliation{3) Theoretical Physics Center for Science Facilities,
Chinese Academy of Sciences, Beijing 100049, P.R. China}

\date{\today}

\begin{abstract}

Intermediate hadronic meson loop contributions to $J/\psi$,
$\psi^\prime \to \gamma\eta_c$ ($\gamma\eta_c^\prime$) are studied
apart from the dominant M1 transitions in an effective Lagrangian
approach. Due to the property of the unique antisymmetric tensor
coupling in $V\to VP$, the hadronic loop transitions provide
explicit corrections to the M1 transition amplitudes derived from
the naive ``quenched" $c\bar{c}$ transitions via the coupling form
factors. This mechanism interfering with the M1 transition
amplitudes naturally accounts for the deviations from the
Godfrey-Isgur model predictions in $J/\psi$ and $\psi^\prime\to
\gamma\eta_c$. It also predicts a small branching ratio of
$\psi^\prime\to \gamma\eta_c^\prime$, which can be examined by
experimental measurements at BES and CLEO-c.

\end{abstract}

\maketitle



\section{Introduction}

Charmonium spectrum and decays of charmonium states are an ideal
place for studying the strong interaction dynamics in the
interplay of perturbative and non-perturbative QCD regime. In the
past decades there have been significant progresses on the
measurement of charmonium spectrum and their decays, which provide
important constraints on phenomenological approaches.

As the first charmonium state discovered in the history, $J/\psi$
has been one of the most widely studied states in both experiment
and theory. As a relatively heavier system compared with light
$q\bar{q}$ mesons, the application of a nonrelativistic potential
model (NR model) including color Coulomb plus linear scalar
potential and spin-spin, spin-orbit interactions, has provided a
reasonably good prescription for the charmonium spectrum. This
success is a direct indication of the validity of the naive
``quenched" $c\bar{c}$ quark model scenario as a leading
approximation in many circumstances. A relativised version was
developed by Godfrey and Isgur~\cite{godfrey-isgur} (GI model),
where a flavor-dependent potential and QCD-motivated running
coupling are employed.  In comparison with the nonrelativistic
model, the GI model offers a reasonably good description of the
spectrum and matrix elements of most of the $u$, $d$, $s$, $c$ and
$b$ quarkonia~\cite{godfrey-isgur,barnes-godfrey-swanson-2005}.

On the other hand, there also arise apparent deviations in the
spectrum observables which give warnings to a simple $q\bar{q}$
treatment and more complicated mechanisms may play a role. As
pointed out in Ref.~\cite{barnes-godfrey-swanson-2005}, the
importance of mixing between quark model $q\bar{q}$ states and two
meson continua may produce significant effects in the spectrum
observables. By including the meson loops, the quark model is
practically ``unquenched". This immediately raises questions about
the range of validity of the naive ``quenched" $q\bar{q}$ quark
model scenario, and the manifestations of the intermediate meson
loops in charmonium spectrum and their decays. These issues become
an interesting topic in the study of charmonium spectrum with
high-statistic charmonium events from experiment. An example is
the newly identified state $X(3872)$ and a possible assignment for
it as a mixture of $c\bar{c}$ and
$D\bar{D^*}$~\cite{tornqvist,swanson}, or open charm
effects~\cite{eichten-2004}.

In the recent years, the intermediate meson loop is investigated
in a lot of meson decay
channels~\cite{li-bugg-zou,cheng,anisovich,zzm,zz-1810,zhao-etac,wzz,li-zhao-zou,liu}
as one of the important non-perturbative transition mechanisms, or
known as final state interactiions (FSI). In particular in the
energy region of charmonium masses, with more and more data from
Belle, BaBar, CLEO-c and BES, it is widely studied that
intermediate meson loop may account for apparent OZI-rule
violations~\cite{zzm,zz-1810,zhao-etac,wzz,li-zhao-zou,liu} via
quark-hadron duality
arguement~\cite{lipkin,geiger-isgur,lipkin-zou}.

In this work, we shall study the radiative decays of $J/\psi$ and
$\psi^\prime$ into $\gamma\eta_c$ and $\gamma\eta_c^\prime$. In
the naive $q\bar{q}$ scenario, this type of decays is dominantly
via magnetic dipole (M1) transitions which flip the quark spin.
For $J/\psi\to \gamma \eta_c$ and $\psi^\prime\to
\gamma\eta_c^\prime$, where the initial and final state $c\bar{c}$
are in the same multiplet, the spatial wavefunction overlap is
unity at leading order, while $\psi^\prime\to\gamma\eta_c$ will
vanish due to the orthogonality between states of different
multiplets. In this sense, the former decays are ``allowed" while
the latter is ``hindered". However, the inclusion of relativistic
corrections from the quark spin-dependent potential will induce a
nonvanishing overlap between states of different multiplets such
that the decay of $\psi^\prime\to \gamma\eta_c$ is
possible~\cite{Eichten:1974af,Eichten:1975ag,Eichten:1978tg,Eichten:1979ms}.

Theoretical studies of the heavy quarkonium M1 transitions with
relativistic corrections are various in the
literature~\cite{Zambetakis:1983te,grotch-84,zhang-91,ebert-2003,lahde-2003,Godfrey:2001vc,brambilla-2006,dudek-2006}.
Relativistic quark model calculations show that a proper choice of
the Lorentz structure of the quark-antiquark interaction in a
meson is crucial for explaining the $J/\psi\to \gamma\eta_c$
data~\cite{ebert-2003}. Systematic investigation of the M1
transitions in the framework of nonrelativistic effective field of
QCD has been reported in Ref.~\cite{brambilla-2006}, where
relativistic corrections of relative order $v^2$ are included. For
$J/\psi\to \gamma\eta_c$ the authors found
$\Gamma_{J/\psi\to\gamma\eta_c}=(1.5\pm 1.0)$ keV, which is in a
good agreement with the data, but with quite large estimated
uncertainties from higher-order relativistic corrections. Taking
into account the transitions of $\psi^\prime\to\gamma\eta_c$, the
overall results for the M1 transitions still turn out to be
puzzling~\cite{eichten-2007,Brambilla:2004wf}. As studied by
Barnes, Godfrey and Swanson~\cite{barnes-godfrey-swanson-2005} in
the NR model and the relativied GI model, although the models
provide an overall consistent description of most of the existing
charmonium states, theoretical results for the M1 transition have
significant discrepancies compared with the experimental
data~\cite{pdg2006}. For example, in both NR and GI model, the
predicted partial decay widths for $J/\psi\to \gamma\eta_c$ are as
large as about two times of the experimental value, while for
$\psi^\prime\to \gamma\eta_c$, the theoretical predictions are
about one order of magnitude larger than the data~\cite{pdg2006}.
For $\psi^\prime\to \gamma\eta_c^\prime$, although the predicted
partial decay widths 0.17-0.21 keV are smaller than the
experimental upper limit ($<0.67$ keV), it is possible that the M1
transition is very different from the experimental measurement.

Therefore, it is likely that there exist additional mechanisms
beyond the $c\bar{c}$ transitions. This consideration thus prompts
us to explore possible sources which can contribute to the
charmonium radiative decay and cause deviations from the NR and GI
model predictions, among which the intermediate meson loop
transitions could be a natural mechanism.

As follows, we first brief the calculations from the NR and GI
models for the M1 transitions, and then introduce the formalisms
for the intermediate meson loop contributions in Section II. The
results and discussions will be presented in Section III.

\section{M1 transition in NR and GI model}

The detailed study of the M1 transition was given by Barnes {\it
et al.} in Ref.~\cite{barnes-godfrey-swanson-2005}, and here we
quote their standard formula to incorporate the intermediate meson
loop contributions which is to be introduced later.

In Ref.~\cite{barnes-godfrey-swanson-2005}, the partial decay
width via M1 transition is evaluated by
\be\label{M1}
\Gamma_{M1}(n^{2S+1}L_J\to
{n^{\prime}}^{2S^\prime+1}L^{\prime}_{J^\prime} +\gamma) = \frac
43\frac{2J^\prime+1}{2L+1}\delta_{LL^\prime}\delta_{S,S^\prime\pm
1} \frac{e^2_c\alpha}{m^2_c}|\langle\psi_f|\psi_i\rangle|^2
E_\gamma^3\frac{E_f}{M_i} \ ,
\ee
where $n$ and $n^\prime$ are the main quantum number of the
initial and final state charmonium meson; $S$ ($S^\prime$), $L$
($L^\prime$) and $J$ ($J^\prime$) are the initial (final) state
spin, orbital angular momentum and total angular momentum.
$E_\gamma$ and $E_f$ denote the final state photon and meson
energy, respectively, while $M_i$ is the initial $c\bar{c}$ meson
mass. $|\psi_i\rangle$ and $|\psi_f\rangle$ are the spatial
wavefunctions of the initial and final state $c\bar{c}$ mesons,
respectively.

In the GI model, phase space factor $E_f/M_i$ is not included
though it is close to unity in many considered cases. In both GI
and NR model, a recoil factor $j_0(kr/2)$ is included. We quote
the results from Ref.~\cite{barnes-godfrey-swanson-2005} for
future comparison.

In order to incorporate the intermediate meson loop contributions,
we derive the effective $V\gamma P$ couplings due to the M1
transition from Eq.~(\ref{M1}) by defining
\be
{\cal M}_{fi}(M1) \equiv \frac{g_{V\gamma
P}}{M_i}\varepsilon_{\alpha\beta\mu\nu} P_\gamma^\alpha
\varepsilon_\gamma^\beta P_i^\mu\varepsilon_i^\nu \ ,
\ee
where $P_i$ and $P_\gamma$ are four-vector momentum of the initial
meson and final state photon, respectively, and $\varepsilon_i$
and $\varepsilon_\gamma$ are the corresponding polarization
vectors. From Ref.~\cite{barnes-godfrey-swanson-2005}, we know
that these extracted effective $g_{V\gamma P}$ couplings for
$J/\psi, \psi^\prime\to \gamma \eta_c$ apparently overestimate the
experimental data. Thus, the introduction of the intermediate
meson loop contributions, which unquench the naive $c\bar{c}$
configurations, is supposed to cancel the M1 transition amplitudes
via destructive interferences.

\section{Intermediate meson loop contributions}

The inclusion of the intermediate meson loops in meson decays
somehow ``unquenches" the naive quark model. A full consideration
of such an effect requires systematic coupled channel calculations
for e.g. the charmonium mass spectrum~\cite{barnes-discussion}. An
interesting feature arising from the low-lying charmonia, such as
$\eta_c$, $\eta_c^\prime$, $J/\psi$, and $\psi^\prime$, is that
their masses are lower than the open charmed meson decay channels.
As a consequence, the lowest open charmed meson decay channels are
expected to be dominant if they indeed account for contributions
beyond the M1 transitions. This scenario turns to be consistent
with the break-down of local quark-hadron duality, where the
leading contributions to the sum over all intermediate virtual
states are from those having less virtualities.

It should be pointed that intermediate states involving flavor
changes turn out to be strongly suppressed. One reason is because
of the large virtualities involved. The other is because of the
OZI rule suppressions. Therefore, intermediate state contributions
such as $\rho\pi$ etc, are negligibly small.

\begin{figure}[ht]
 \begin{center}
\epsfig{file=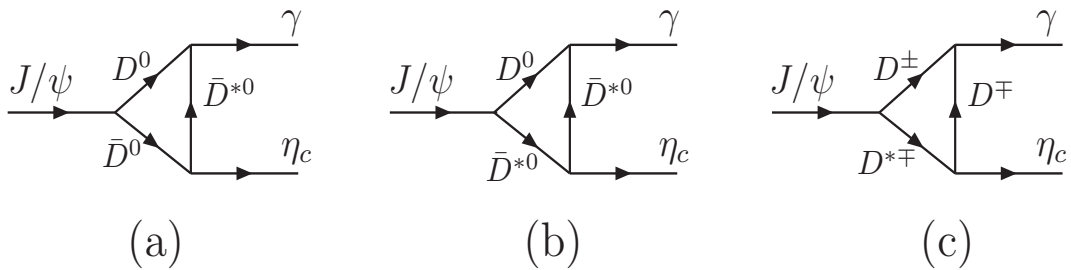, width=14cm,height=3.5cm}
\caption{Schematic diagrams for  $J/\psi\to \gamma\eta_c$ via (a)
$D\bar{D}(D^*)$, (b) $D\bar{D^*}(D^*)$ and (c) $D\bar{D^*}(D)$
intermediate meson loops. Similar processes occur in
$\psi^\prime\to \gamma\eta_c$ and $\gamma\eta_c^\prime$. }
 \protect\label{fig-1}
 \end{center}
\end{figure}

Following the above consideration, we thus investigate
$D\bar{D}(D^*)$, $D\bar{D^*}(D^*)$ and $D\bar{D^*}(D)$ loops as the
major contributions to $J/\psi\to \gamma\eta_c$, and
$\psi^\prime\to\gamma\eta_c, \ \gamma\eta_c^\prime$ as illustrated
in Fig.~\ref{fig-1}. We stress that although some of the vertices in
the loop may violate gauge invariance, such as $J/\psi D\bar{D}$,
the overall antisymmetric property is retained for the loops. The
loop contributions hence only provide corrections to the $VVP$
coupling strength for the external fields, but not change their
antisymmetric tensor structure, no matter $V$ is a massive vector
meson or photon. Apart from the transitions in Fig.~\ref{fig-1}, the
contact transitions in Fig.~\ref{fig-2} will also contribute to the
decay amplitude. We show that the processes of Fig.~\ref{fig-2} are
gauge invariant by themselves. In brief, due to the property of the
antisymmetric tensor coupling of $VVP$, where both $V$ and $P$ are
external fields here, hadronic loop corrections are guaranteed to be
gauge invariant in this effective Lagrangian approach.

The detailed formulation is given in the following subsections.

\subsection{Intermediate $D\bar{D}(D^*)+c.c.$ loop}

The transition amplitude for an initial vector charmonium
($J/\psi$ or $\psi^\prime$) decay into $\gamma \eta_c$ or
$\gamma\eta_c^\prime$ via $D\bar{D}(D^*)$ can be expressed as
follows:
\be\label{dd-loop}
{\cal M}_{fi}=\int \frac {d^4p_2}{(2\pi)^4}\sum_{D^* pol}
 \frac {T_1T_2T_3}{a_1a_2a_3}{\cal F}(p_2^2) \ ,
\ee
where the vertex functions are
\be\label{vertex-1}
\left\{\begin{array}{ccc}
 T_1 &\equiv& ig_1(p_1-p_3)\cdot \varepsilon_i \\
 T_2&\equiv& \frac
 {ig_2}{m_2}\varepsilon_{\alpha\beta\mu\nu}P_\gamma^\alpha\varepsilon_\gamma^\beta
 p_2^\mu\varepsilon_2^\nu \\
 T_3&\equiv& ig_3(P_f+p_3)\cdot \varepsilon_2\end{array}\right.
 \ee
where $g_1$, $g_2$, and $g_3$ are the coupling constants at the
meson interaction vertices (see Fig. \ref{fig-1}). The four
vectors, $P_i$, $P_\gamma$, and $P_f$ are the momenta for the
initial vector, final state $\gamma$ and pseudoscalar meson,
respectively, while four-vector momenta, $p_1$, $p_2$, and $p_3$
are for the intermediate mesons, respectively, and
$a_1=p_1^2-m_1^2, a_2=p_2^2-m_2^2$, and $a_3=p_3^2-m_3^2$ are the
denominators of the propagators of these intermediate mesons.

As being studied in Ref.~\cite{li-zhao-zou}, this loop diverges
logarithmically. Thus, a form factor to suppress the divergence
and take into account the momentum-dependence of the vertex
couplings is included:
\bea
{\cal F}(p^2) = \left(\frac {\Lambda^2 - m^2_2}
{\Lambda^2-p_2^2}\right)^n,
\eea
where $n=1,2$ correspond to monopole and dipole form factors,
respectively. An empirical argument applied here is that in the
$P$-wave  $V\to VP$ decay the form factor favors a dipole form. We
hence deduce the loop transition amplitudes with a dipole form
factor.

Substitute the vertex couplings of Eq.~(\ref{vertex-1}) into
Eq.~(\ref{dd-loop}), the integral has an expression:
\be
{\cal M}_{fi}  = \int \frac {d^4p_{2}} {(2\pi)^4}
\sum _{D^* pol } \frac {[ig_1 (p_1 - p_3 ) \cdot
 \varepsilon_i] [\frac
 {ig_2}{m_2}\varepsilon_{\alpha\beta\mu\nu}P_\gamma^\alpha\varepsilon_\gamma^\beta
 p_2^\mu \varepsilon_2^\nu] [ig_3(P_f + p_3)\cdot
 \varepsilon_2]}{(p_1^2-m_1^2)(p_3^2-m_3^2)(p_2^2-m_3^2)}
{\cal F}(p_2^2) \ .
\ee
With a dipole form factor, we have
\be\label{dipole-loop-1}
 {\cal M}_{fi} \equiv \frac{\tilde{g}_a}{M_i} \varepsilon_{\alpha\beta\mu\nu}
 P_\gamma^\alpha\varepsilon_\gamma^\beta P_i^\mu\varepsilon_i^\nu
 \ ,
\ee
where
\bea
 \tilde{g}_a &\equiv & -\frac {g_1 g_2 g_3 M_i}{m_2}
 \int^{1}_{0}dx\int^{1-x}_{0}dy
 \frac{2}{(4\pi)^2}\left[\log \frac {\triangle (m_1,m_3,\Lambda)}{\triangle (m_1,m_3,m_2)}\right.  \nonumber\\
 & &\left.- \frac {y(\Lambda^2 - m_2^2)}{\triangle (m_1,m_3,\Lambda
 ) }\right]
\eea where the function $\Delta$ is defined as \bea
  \Delta(a,b,c)\equiv  -(M_i^2-M_f^2)(1-x-y)x+M_f^2 x^2
 +a^{2}(1-x-y) -(M_f^{2}-b^{2})x+y c^2  \ .
\eea

In the intermediate meson exchange loop, coupling $g_2$ can be
determined via the experimental information for $ D^{* 0}\to
D^0\gamma({\bar D}^{* 0}\to {\bar D}^0\gamma) $, i.e.
\be
g_2^2=\frac {12\pi M_{D^*}^2 } {|p_\gamma|^3}\Gamma_{D^{* 0}\to
D^0\gamma},
\ee
where $\Gamma_{D^{* 0}\to D^0\gamma} = (38.1\pm 2.9)\% \times
\Gamma_{tot}$ is given by experiment~\cite{pdg2006}. We neglect
the contributions from the charged meson exchange loop since
$\Gamma_{D^{*\pm}\to D^\pm\gamma}=(1.6\pm 0.4)\%\times 96$ keV is
about two orders of magnitude smaller than $\Gamma_{D^{* 0}\to
D^0\gamma}$.

For coupling constant $g_1$, especially $g_{J/\psi D\bar{D}}$,
there are several methods suggested in the literature including
quark model using heavy quark effective theory
approach~\cite{Deandrea:2003pv}, QCD sum
rule~\cite{Matheus:2002nq,Bracco:2004rx},  SU(4) symmetry and
vector meson dominance (VMD) model~\cite{Lin:1999ad}. They
typically give a value of order of one for $g_{J/\psi D\bar{D}}$.
In this work, we adopt $g_{J/\psi D\bar{D}} =7.20$ which is
consistent with the value from Ref.~\cite{Deandrea:2003pv}.

For the $g_{D^* D \eta_c}$ coupling, we assume
\be
 g_{D^{*0}D^0\eta_c}= g_{J/\psi D^{0*} \bar{D}^0}  \ .
\ee

\subsection{Intermediate $D\bar{D^*}(D^*)+c.c.$ loop}

As shown by Fig.~\ref{fig-1}(b), the transition amplitude from the
intermediate $D\bar{D^*}(D^*)+c.c.$ loop can be expressed the same
form as Eq.~(\ref{dd-loop}) except that the vertex functions
change to
\be\label{vertex-1b}
\left\{\begin{array}{ccl}
 T_1 &\equiv &\frac {i f_1}{M_i}
 \varepsilon_{\alpha\beta\mu\nu}
 P_i^\alpha \varepsilon_i^\beta p_3^\mu \varepsilon_3^\nu \ , \\
 \nonumber
 T_2&\equiv& \frac {i f_2}{m_2}
 \varepsilon_{\alpha^\prime\beta^\prime\mu^\prime\nu^\prime}
 p_2^{\alpha^\prime} \varepsilon_2^{\beta^\prime} P_\gamma^{\mu^\prime} \varepsilon_\gamma^{\nu^\prime} \ , \\
 T_3&\equiv & \frac {i f_3}{M_f}
 \varepsilon_{\alpha^{\prime\prime}\beta^{\prime\prime}\mu^{\prime\prime}\nu^{\prime\prime}}
 p_2^{\alpha^{\prime\prime}} \varepsilon_2^{\beta^{\prime\prime}} p_3^{\mu^{\prime\prime}} \varepsilon_3^{\nu^{\prime\prime}}
\end{array}\right.
 \ee
where $f_{1,2,3}$ are the coupling constants. With a dipole form
factor the integration gives
\be\label{dipole-loop-2b}
 {\cal M}_{fi}\equiv \frac{\tilde{g}_b}{M_i}
\varepsilon_{\alpha\beta\mu\nu}
 P_\gamma^\alpha\varepsilon_\gamma^\beta P_i^\mu\varepsilon_i^\nu
 \ ,
\ee where \bea \tilde{g}_b &\equiv & \frac {f_1 f_2
f_3}{m_2M_f}
 \int^{1}_{0}dx\int^{1}_{0}dy\int^{1-x-y}_0 dz(1-x-y-z)
 \frac{2}{(4\pi)^2}( \frac {A}{\triangle_1^2}
 - \frac {B}{\triangle_1^3}) \ ,
\eea
with
\bea A &=&\frac {1}{4}[(1-x-\frac{z}{2})(M_i^2-M_f^2)+xM_f^2] \ , \nonumber\\
B &=&
-\frac{x}{4}(M_i^2-M_f^2)[z(1-x)(M_i^2-3M_f^2)-(M_i^2-M_f^2)] \ ,
\nonumber\\
\Delta_1 &=& -x z(M_i^2-M_f^2) +z m_1^2 +y m_2^2 +x m_3^2
+(1-x-y-z) \Lambda^2 \ .
\eea

In the above equation the intermediate meson masses $m_{1,2,3}$
are from the $D\bar{D^*}(D^*)$ loop, which are different from
those in Eq.~(\ref{dipole-loop-1}). $f_{1,2,3}$ denotes the
corresponding vertex coupling constants.

In the $D\bar{D^*}(D^*)+c.c.$ loop, the coupling constant
$g_{J/\psi D^* \bar{D}}$ is related to $g_{J/\psi D\bar{D}}$ by
the relation of the heavy quark mass limit~\cite{Deandrea:2003pv}:
\bea
g_{J/\psi D^*\bar{D}} = g_{J/\psi D\bar{D}}/\tilde{M}_D ,
\eea
where $\tilde{M}_D$ corresponds to the mass ratio of
$M_D/M_{D^*}$. Similarly, we have $g_{\psi^\prime
D^*\bar{D}}=g_{\psi^\prime D\bar{D}}/\tilde{M}_D$.

For $\psi^\prime\to \gamma\eta_c$ and $\gamma\eta_c^\prime$, we
assume that $g_{\psi^\prime D\bar{D}}=g_{J/\psi D\bar{D}}$,
${g_{D^{*0}D^0\eta_c}=g_{J/\psi D^*\bar{D}}}$, and
${g_{D^{*0}D^0\eta_c^\prime}=g_{\psi^\prime D^\ast \bar{D}} }$,
which are consistent with the $^3 P_0$
model~\cite{barnes-godfrey-swanson-2005}. In Table~\ref{tab-1} the
values of the coupling constants are listed.

\subsection{Intermediate $D\bar{D^*}(D)+c.c.$ loop}

The transition amplitude from the intermediate
$D\bar{D^*}(D)+c.c.$ loop can contribute via charged intermediate
meson exchange. Treating the intermediate mesons as fundamental
degrees of freedom, we eventually neglect the contributions from
the non-zero magnetic moments of the $D$ mesons. The
charge-neutral loop is thus suppressed due to the vanishing
$D^0\bar{D^0}\gamma$ electric coupling. Therefore, we only
consider the charged meson loop contributions as shown by
Fig.~\ref{fig-1}(c). The transition amplitude can also be
expressed in a form as Eq.~(\ref{dd-loop}) with the vertex
functions
\be\label{vertex-2}
 \left \{\begin{array}{ccl}
 T_1 &\equiv &\frac {i f_1^\prime}{M_i}
 \varepsilon_{\alpha\beta\mu\nu}
 P_i^\alpha \varepsilon_i^\beta p_3^\mu \varepsilon_3^\nu \ , \\
 T_2&\equiv& i f_2^\prime (p_1-p_2)\cdot \varepsilon_\gamma \ , \\
 T_3&\equiv & i f_3^\prime (P_f-p_2)\cdot \varepsilon_3 \ , \end{array}
 \right.
\ee
where $f^\prime_{1,2,3}$ are the coupling constants and and ${\cal
F}(p_2^2)$ is the form factor. With a dipole form factor the
integration gives
\be\label{dipole-loop-2}
 {\cal M}_{fi}\equiv \frac{\tilde{g}_c}{M_i}
\varepsilon_{\alpha\beta\mu\nu}
 P_\gamma^\alpha\varepsilon_\gamma^\beta P_i^\mu\varepsilon_i^\nu
 \ ,
\ee
where
\bea
\tilde{g}_c &\equiv & f_1^\prime f_2^\prime f_3^\prime
 \int^{1}_{0}dx\int^{1-x}_{0}dy
 \frac{2}{(4\pi)^2}\left[\log \frac {\triangle (m_1,m_3,\Lambda)}{\triangle (m_1,m_3,m_2)}\right.  \nonumber\\
 & &\left.- \frac {y (\Lambda^2 - m_2^2)}{\triangle (m_1,m_3,\Lambda
 ) }\right] \ .
\eea

It is interesting to note that the integral of the
$D\bar{D^*}(D)+c.c.$ loop has a similar form as that of
$D\bar{D}(D^*)$. However, we expect that contributions from this
loop integral will be relatively suppressed since coupling
$f_2^\prime$ is taken as the unit charge $e =
(4\pi\alpha_e)^{1/2}$.

\subsection{Contact diagrams}

\begin{figure}
\begin{center}
\epsfig{file=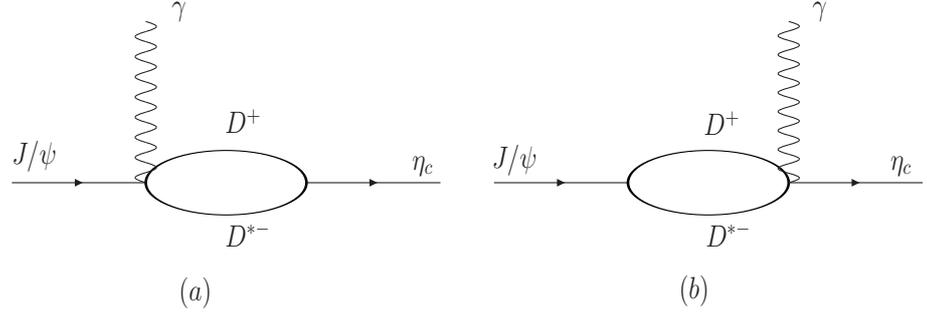, width=14cm,height=24cm}
\vspace{-17.0cm}\caption{The contact diagrams considered in
$J/\psi\to\gamma\eta_c $. Similar diagrams are also considered in
$\psi^\prime\to\gamma\eta_c (\eta_c^\prime)$.}\protect\label{fig-2}
\end{center}

\end{figure}

The contact diagrams of Fig.~\ref{fig-2}(a) and (b) (as an example
in $J/\psi\to \gamma\eta_c$) arise from gauging the strong
$J/\psi(\psi^\prime) D^\ast D$ and $\eta_c (\eta_c^\prime) D^\ast D$
interaction Lagrangians containing derivatives. The general form of
the transition amplitude of Fig.~\ref{fig-2}(a) and (b) can be
expressed as follows:
\be\label{contact-loop} {\cal M}_{fi}=\int \frac
{d^4p_2}{(2\pi)^4}\sum_{D^* pol}
 \frac {T_1T_2}{a_1a_2}{\cal F}(p_2^2) \ ,
\ee
 where ${\cal F}(p_2^2)$ is the form factor as before,
and $T_i(i=1,2)$ are the vertex functions. For Fig. \ref{fig-2}(a),
the expressions of $T_i(i=1,2)$ are:
\be\label{vertex-4}
 \left \{\begin{array}{ccl}
 T_1 &\equiv &\frac {h_1 e}{M_i}
 \varepsilon_{\alpha\beta\mu\nu}
 (\varepsilon_\gamma^\alpha \varepsilon_i^\beta p_2^\mu \varepsilon_2^\nu +P_i^\alpha \varepsilon_i^\beta \varepsilon_\gamma^\mu \varepsilon_2^\nu)\ , \\
 T_2&\equiv& 2i  h_2 P_f\cdot \varepsilon_2 \ , \end{array}
 \right.
\ee where $h_{1,2}$ represent the $J/\psi (\psi^\prime) D^\ast
\bar{D}$ and $\eta_c (\eta_c^\prime) D^\ast \bar{D}$ coupling
constants, respectively, and their values have been given in
Table~\ref{tab-1}.

Using the Feynman parameter scheme, the amplitude for Fig.
\ref{fig-2}(a) can be reduced to
 \bea
\nonumber M_{fi} &=& \frac {2ih_1 h_2
e}{M_i}\varepsilon_{\alpha\beta\mu\nu}\int \frac {d^4p_2}{(2\pi)^4}
\frac {[\varepsilon_\gamma^\alpha \varepsilon_i^\beta p_2^\mu
p_f^\nu +
P_i^\alpha\varepsilon_i^\beta\varepsilon_\gamma^\mu(-P_f^\nu+\frac
{P_f^\rho p_2^\rho p_2^\nu} {m_2^2})
](m_2^2-\Lambda^2)^2}{(p_1^2-m_1^2)(p_2^2-m_2^2)(p_2^2-\Lambda^2)^2}
\\  &=& \frac {2ih_1 h_2
e}{M_i}\varepsilon_{\alpha\beta\mu\nu}\int \frac {d^4l}{(2\pi)^4}
\frac {[\varepsilon_\gamma^\alpha \varepsilon_i^\beta (l-xP_f)^\mu
p_f^\nu +
P_i^\alpha\varepsilon_i^\beta\varepsilon_\gamma^\mu(-P_f^\nu+\frac
{P_{f\rho} (l-xP_f)^\rho (l-xP_f)^\nu} {m_2^2})
](m_2^2-\Lambda^2)^2}{(l^2-\triangle_2)^4}\nonumber\\
& = &\frac{h_1 h_2 e}{M_i} \varepsilon_{\alpha\beta\mu\nu}
 P_\gamma^\alpha\varepsilon_\gamma^\beta P_i^\mu\varepsilon_i^\nu\int_0^1 dx\int_0^{1-x} dy \frac {1}
{(4\pi)^2} (\frac {1} {3\triangle_2^2} - \frac {1} {6\triangle_2})
\\
&\equiv & \frac{\tilde{g}_d}{M_i} \varepsilon_{\alpha\beta\mu\nu}
 P_\gamma^\alpha\varepsilon_\gamma^\beta P_i^\mu\varepsilon_i^\nu\ ,
 \label{contact-loop-int}\eea
with
 \be \triangle_2 = x^2M_f^2
-xM_f^2+xm_1^2+ym_2^2+(1-x-y)\Lambda^2 \ ,\ee
 and
\be {\tilde{g}_d}\equiv  h_1 h_2 e\int_0^1 dx\int_0^{1-x} dy \frac
{1} {(4\pi)^2} (\frac {1} {3\triangle_2^2} - \frac {1}
{6\triangle_2}).\ee

For Fig.~\ref{fig-2}(b), the vertex functions are:
\be\label{vertex-5}
 \left \{\begin{array}{ccl}
 T_1 &\equiv &\frac {i h_1 }{M_i}
 \varepsilon_{\alpha\beta\mu\nu}
 P_i^\alpha \varepsilon_i^\beta p_2^\mu \varepsilon_2^\nu \ , \\
 T_2&\equiv& 2e  h_2 \varepsilon_\gamma \cdot \varepsilon_2 \ , \end{array}
 \right.
\ee
 The amplitude can then be reduced to
 \bea M_{fi} &=& \frac {-2ih_1 h_2
e}{M_i}\varepsilon_{\alpha\beta\mu\nu}P_i^\alpha \varepsilon_i^\beta
\varepsilon_\gamma^\nu \int \frac {d^4p_2}{(2\pi)^4}\frac {p_2^\mu}
{(p_1^2-m_1^2)(p_2^2-m_2^2)(p_2^2-\Lambda^2)^2} \\
&=& \frac {-2ih_1 h_2
e}{M_i}\varepsilon_{\alpha\beta\mu\nu}P_i^\alpha \varepsilon_i^\beta
\varepsilon_\gamma^\nu \int \frac {d^4 l}{(2\pi)^4}\frac
{(l-xP_i)^\mu} {(l^2-\triangle_3)^4} , \eea
 with
 \be \triangle_3 =
x^2M_i^2 -xM_i^2+xm_1^2+ym_2^2+(1-x-y)\Lambda^2.\ee
 Note that the integrand has an odd power of the internal momentum,
 the amplitude will vanish and has no contribution to the $VVP$ coupling.

The above deduction shows that only Fig.~\ref{fig-2}(a) has
nonvanishing contributions to the transition amplitude. Meanwhile,
gauge invariance is also guaranteed for the contact diagrams. The
divergence of the loop integral is eliminated by adding the dipole
form factor as in Fig.~\ref{fig-1}.

\section{Results and discussions}

Proceed to numerical results from the intermediate meson exchange
loops, the undetermined quantities include the cut-off energy
$\Lambda$ in the dipole form factor and the relative phases among
those amplitudes.  The transition amplitude accommodating the M1 and
intermediate meson exchange loops, i.e. $D\bar{D}(D^*)$,
$D\bar{D^*}(D^*)$, $D\bar{D^*}(D)$, and the contact term, can then
be expressed as
 \be {\cal M}_{fi}=\frac{1}{M_i}[g_{V\gamma P}+
\tilde{g}_a e^{i\delta_a}+  \tilde{g}_b e^{i\delta_b} + \tilde{g}_c
e^{i\delta_c}+ \tilde{g}_d e^{i\delta_d}]
\varepsilon_{\alpha\beta\mu\nu}
 P_\gamma^\alpha\varepsilon_\gamma^\beta P_i^\mu\varepsilon_i^\nu
 \ ,
\ee where $g_{V\gamma P}$ is a real number and fixed to be positive.
Couplings $\tilde{g}_a$, $\tilde{g}_b$, $\tilde{g}_c$ and
$\tilde{g}_d$, calculated by the loop integrals can be complex
numbers in principle. In this interested case, since the decay
threshold of the intermediate mesons are above the initial meson
($J/\psi$ and $\psi^\prime$) masses, the absorptive part of the loop
integrals vanishes as a consequence. However, there might exist
relative phases among those transition amplitudes. We hence include
possible relative phases $\delta_{a,b,c,d}$ in the above expression.

Note that the loop contributions are supposed to provide
cancellations to the M1 amplitude which is real. We thus simply take
$\delta_{a,b,c,d}=0$ or $\pi$. In this way, we have several phase
combinations which are to be examined in the numerical calculation.
Yet there is still a free parameter $\Lambda$ to be constrained.

We find that a reasonable constraint on the model can be achieved
by requiring a satisfactory of the following conditions: i) For
either constructive ($\delta=0$) or destructive phases
($\delta=\pi$), the same value of $\Lambda$ is needed to account
for $J/\psi\to \gamma\eta_c$, $\psi^\prime\to \gamma\eta_c$
simultaneously. ii) The value of $\Lambda$ is within the commonly
accepted region, $1.5\sim 2.5$ GeV. iii) The prediction for
$\psi^\prime\to\gamma\eta_c^\prime$ with the same $\Lambda$ is
well below the experimental upper limit,
$BR(\psi^\prime\to\gamma\eta_c^\prime)<2.0\times
10^{-3}$~\cite{pdg2006}.

Imposing the above conditions on fitting the $\Lambda$ parameter, we
obtain $\Lambda=2.39$ GeV as the best fit with
$\delta_{a,b,c,d}=\pi$, i.e. contributions from the loop integrals
provide cancellations to the M1 transition amplitudes and there is
no need for abnormal relative phases among the intermediate meson
exchanges.

The numerical results for the intermediate meson exchanges have some
predominant features. We find that the $D\bar{D}(D^*)$ and
$D\bar{D^*}(D^*)$ loops have relatively large contributions while
the $D\bar{D^*}(D)$ loop is quite small. The contributions from the
contact term are negligibly small in $J/\psi\to\gamma\eta_c$ and
$\psi^\prime\to\gamma\eta_c^\prime$, while relatively large in
$\psi^\prime\to\gamma\eta_c$.

In Table~\ref{tab-2}, the fitted branching ratios are listed and
compared with the GI model M1 transitions. We also list the
exclusive contributions from the triangle diagrams of
Fig.~\ref{fig-1} and contact diagrams of Fig.~\ref{fig-2} as a
comparison. For $J/\psi\to \gamma\eta_c$, we find that the magnitude
of the meson loop amplitude is smaller than the M1 amplitude, while
for $\psi^\prime\to\gamma\eta_c$, the absolute loop amplitude turns
to be larger than the M1. With $\Lambda=2.39$ GeV, we obtain
$\Gamma(J/\psi\to\gamma\eta_c)=1.59$ keV which is located at the
upper limit of the experimental data,
$\Gamma_{exp}(J/\psi\to\gamma\eta_c)=(1.21\pm 0.37)$
keV~\cite{pdg2006}. For $\psi^\prime\to\gamma\eta_c$, we have
$\Gamma(\psi^\prime\to\gamma\eta_c)=0.86$ keV, which is agree well
with the data, $\Gamma_{exp}(\psi^\prime\to\gamma\eta_c)=(0.88\pm
0.13)$ keV~\cite{pdg2006}. Taking into account the still-large
uncertainties with the data for $J/\psi\to \gamma\eta_c$, the
inclusion of the intermediate meson loop contributions significantly
improves the theoretical results.

With the fixed $\Lambda$, the partial decay width for
$\psi^\prime\to\gamma\eta_c^\prime$ is calculated as a prediction.
The pure M1 transition predicts
$\Gamma^{GI}_{M1}(\psi^\prime\to\gamma\eta_c^\prime)\simeq 0.17$
keV~\cite{barnes-godfrey-swanson-2005}, while the hadronic loops
contribute $\Gamma_{HL}(\psi^\prime\to\gamma\eta_c^\prime)\simeq
0.054$ keV. The cancellation from the hadron loops thus leads to
$\Gamma_{all}(\psi^\prime\to\gamma\eta_c^\prime)\simeq 0.032$ keV
which is well below the experimental upper limit, 0.67
keV~\cite{pdg2006}. Note that in all these three channels, the
hadronic loop cancellations from the real part of the amplitudes
possess the same relative sign to the M1 amplitudes. This makes the
decay of $\psi^\prime\to\gamma\eta_c^\prime$ extremely interesting.
As the pure M1 transition still predicts a sizeable partial width
about 0.17 keV while our hadronic loop cancellation predicts a much
smaller value, improved measurement of this quantity will help us
gain further insights into the decay mechanisms.

For other relative phases, we find that there does not exist a
common value for $\Lambda$ to fit the data for
$J/\psi\to\gamma\eta_c$ and $\psi^\prime\to \gamma\eta_c$
simultaneously.

To summarize, in this work we have studied the hadronic meson loop
contributions to the $J/\psi$ and $\psi^\prime$ radiative decays
into $\gamma\eta_c$ or $\gamma\eta_c^\prime$. In the framework of
effective Lagrangian phenomenology, the intermediate meson
exchange loops provide corrections to the leading couplings
extracted from potential quark models. In comparison with the NR
and GI model, the meson loop contributions turn to cancel the NR
and GI amplitudes. It is interesting to see that the meson loop
contributions in $J/\psi\to\gamma\eta_c$ is smaller than the M1
transition in magnitude, while in $\psi^\prime\to \gamma\eta_c$
the situation is opposite. Note that the pure M1 contribution in
$\psi^\prime\to \gamma\eta_c$ is about one order of magnitude
larger than the experimental data, the meson loop contributions
turn out to be even larger. This mechanism suggests significant
cancellations between the M1 and meson loop amplitudes. It raises
questions on the naive $q\bar{q}$ solution for the meson spectrum,
and could be a manifestation of the limit of the quenched quark
model scenario.

As a prediction from this model, we calculate the partial decay
width of $\psi^\prime\to \gamma\eta_c^\prime$. It gives a value
about one order of magnitude smaller than the experimental upper
limit. Improved measurement of this decay channel is strongly
recommended.

It is interesting to note that our model results are similar to
those from a relativistic quark model calculation by Ebert, Faustov
and Galkin~\cite{ebert-2003}, who find that a proper choice of the
Lorentz structure of the quark-antiquark interaction in a meson is
crucial for accounting for the M1 transition data. In our approach
we extract the effective couplings from the NR and relativised GI
model and then combine it with the gauge invariant meson loop
corrections. The validity of this approach is guaranteed by the
property of the unique antisymmetric tensor coupling for $VVP$
fields. In the framework of effective Lagrangian phenomenology the
corrections to the leading contributions are introduced as coupling
form factors.

Although we also observe strong sensitivities of the hadronic loop
contributions to the cut-off energy $\Lambda$, the advantage of
this approach is that the number of parameters is limited. In
fact, there is little freedom for the effective couplings at
vertices. By a coherent study of $J/\psi$ and
$\psi^\prime\to\gamma\eta_c$, we find that the constraint on the
$\Lambda$ value is very tight. Certainly, it should be noted that
our treatment of the relative phases is empirical though the favor
of a destructive phase between the M1 transition amplitudes and
hadron loops turns to be consistent with what one naturally
expects. Note that it has been shown in Ref.~\cite{Eichten:1979ms}
that a proper modification of the color Coulomb potential strength
will simultaneously account for the branching ratios for
$J/\psi\to \gamma\eta_c$ and $\psi^\prime\to \gamma\eta_c$. It
seems to support that the intermediate hadronic meson loops are
responsible, at least partly, for such a modification, and hence
break down the naive $q\bar{q}$ scenario.

The study of non-perturbative effects arising from intermediate
meson loops in heavy quarkonium decays has attracted a lot of
attention recently. Although such approaches still experience
large uncertainties from the divergent behavior of the loop
integrals, we expect that improved experimental measurements with
high statistics, such as at BES and CLEO-c, provide more and more
stringent constraints on the hadronic loops. Thus, insights into
the effective degrees of freedom within hadrons and their decay
mechanisms can be gained. This requires systematic analysis of
both spectroscopy and coupled channels for which more and more
theoretical efforts are undergoing.

\section*{Acknowledgement}

Q.Z. would like to thank T. Barnes, K.T. Chao, Y. Jia and B.S. Zou
for very useful discussions. This work is supported, in part, by the
U.K. EPSRC (Grant No. GR/S99433/01), National Natural Science
Foundation of China (Grant No.10675131 and 10491306), and Chinese
Academy of Sciences (KJCX3-SYW-N2).


\begin{table}[ht]
\begin{tabular}{|c|c|c|c|c|c|c|c|}
 \hline  Coupling constants & $|g_{J/\psi D\bar{D}}|$  &$|g_{\psi^\prime D\bar{D}}|$   & $|g_{J/\psi {\bar D}^\ast D}|$ &
 $|g_{D^*\bar{D}\eta_c}|$ & $|g_{\psi^\prime {\bar D}^\ast D}|$ & $|g_{D^*\bar{D}\eta_c^\prime}|$ & $|g_{D^*D\gamma}|$ \\ [1ex] \hline
 numerical value
 & 7.20 & 7.20  &4.34 &4.34 &3.64 &3.64 & 6.86\\ [1ex] \hline
\end{tabular}
\caption{ The absolute values of coupling constants for the
effective vertex interactions. Their relative phases are
determined by the SU(4) flavor symmetry.  } \label{tab-1}
\end{table}

\begin{table}
\begin{tabular}{|c|c|c|c|}
 \hline
 Initial meson                 & $J/\psi(1^3S_1)$   &\multicolumn{2}{c|}      {$\psi^\prime(2^3S_1)$}       \\
 \hline
 Final meson                   & $\eta_c(1^1S_0)$   & $\eta_c^\prime(2^1S_0)$   &  $\eta_c(1^1S_0)$        \\
 \hline
 $\Gamma^{NR}_{M1}$ (keV)      & 2.9                & 0.21                      &  9.7                     \\
 \hline
 $\Gamma^{GI}_{M1}$ (keV)      & 2.4                & 0.17                      &  9.6                     \\
 \hline
 $\Gamma_C$         (keV)      & $\sim 0$           & $\sim 0$                  & 0.04                     \\
 \hline
 $\Gamma_{Tri}$     (keV)      & 0.096              & 0.063                     & 17.91                    \\
 \hline
 $\Gamma_{HL}$      (keV)      & 0.083              & 0.054                     & 16.20                    \\
 \hline
 $\Gamma_{all}$     (keV)      & 1.59               & 0.032                     & 0.86                     \\
 \hline
 $\Gamma_{exp}$     (keV)      & $1.21\pm 0.37$     & $< 0.67$                  & $0.88\pm 0.13$           \\
 \hline
\end{tabular}
\caption{ Radiative partial decay widths  given by different
processes are listed: $\Gamma^{NR}_{M1}$ and $\Gamma^{GI}_{M1}$ are
the M1 transitions in the NR and GI model,
respectively~\protect\cite{barnes-godfrey-swanson-2005};
$\Gamma_{Tri}$ are inclusive contributions from the triangle
diagrams (Fig.~\protect\ref{fig-1}); $\Gamma_C$ are from contact
diagrams (Fig.~\protect\ref{fig-2}); $\Gamma_{HL}$ denote the
inclusive contributions from all the intermediate hadronic loops;
while $\Gamma_{all}$ are coherent results including the M1 in the GI
model and intermediate hadronic loops. The experimental data are
from PDG2006~\protect\cite{pdg2006}. The results are obtained at the
cut-off energy $\Lambda=2.39$ GeV.} \protect\label{tab-2}
\end{table}

\end{document}